\documentclass[aps,pre,epsfigure,twocolumn,superscriptaddress]{revtex4-1}
\usepackage[colorlinks=true,linkcolor=blue,urlcolor=blue,citecolor=blue,pdfusetitle]{hyperref}
\usepackage[utf8]{inputenc}
\usepackage[english]{babel}
\usepackage{amsmath}
\usepackage[caption = false]{subfig}
\usepackage{graphicx,epstopdf}
\usepackage{blindtext}
\usepackage[table,xcdraw]{xcolor}
\usepackage{lipsum}
\usepackage{amsfonts}
\usepackage{bbm}
\usepackage{amssymb}
\usepackage{enumerate}
\usepackage{color}
\usepackage{latexsym}
\usepackage{physics}
\usepackage{times,txfonts}
\usepackage{ulem}

\begin{document}

\title{Entanglement dynamics of a dc SQUID interacting with a single mode radiation field}

\author{Cleidson Castro}\email{ccastro@ufrb.edu.br}
\affiliation {\it Centro de Forma\c c\~ao de Professores, Universidade Federal do Rec\^oncavo da Bahia,Avenida Nestor de Mello Pita, 535 Amargosa, Bahia, Brazil.}
\author{Matheus R. Ara\'{u}jo}
\affiliation {\it Grupo de Informa\c{c}\~{a}o Qu\^{a}ntica e F\'{i}sica Estat\'{i}stica, Centro das Ci\^{e}ncias Exatas e das Tecnologias, Universidade Federal do Oeste da Bahia. Rua Bertioga, 892, Morada Nobre I, 47810-059 Barreiras, Bahia, Brazil.}
\author{Clebson Cruz}\email{clebson.cruz@ufob.edu.br}
\affiliation {\it Grupo de Informa\c{c}\~{a}o Qu\^{a}ntica e F\'{i}sica Estat\'{i}stica, Centro das Ci\^{e}ncias Exatas e das Tecnologias, Universidade Federal do Oeste da Bahia. Rua Bertioga, 892, Morada Nobre I, 47810-059 Barreiras, Bahia, Brazil.}

\date{\today}

\begin{abstract} 
In this work, we study the coupling between a superconducting device as a dc SQUID, simulated from an artificial atom with two degrees of freedom, and a single-mode radiation field for the information transference process. We demonstrate that the population transfer among the energy levels of the artificial atom yields an entanglement dynamics, which leads to the generation of a pair of photons. Moreover, we show the quantum information transference between the internal modes of the superconducting device, initially in a maximally entangled state, and the radiation field. The artificial atom absorbs the photon, and the radiation field modes become entangled as quantum coherence is transmitted from the superconducting device to the photons. These results strengthen the applicability of superconducting devices for the transference of quantum information, contributing to promising applications in emerging quantum technologies.
\end{abstract}


\maketitle 

\section{Introduction}
Superconducting circuits (SC) have been received considerable attention in the past few years \cite{Kjaergaard:2020, AppliesPhysics:2019,ReportsProgPhys80:2017,strambini2020josephson,santos2019stable,pal2019quantized,ReportsProgPhys80:2017,Devoret1169} due to their promising applications in emergent quantum technologies, such as quantum computers \cite{Kjaergaard:2020, AppliesPhysics:2019,ReportsProgPhys80:2017} and quantum batteries \cite{strambini2020josephson,santos2019stable,pal2019quantized}. These systems have the advantage of presenting a series of parameters and physical properties that can be controlled accordingly to their intended use~\cite{Quantum_Inf_Process:2009,gu2017microwave,ReportsProgPhys80:2017,Devoret1169}. In this regard, several prototypes have been proposed showing the computational power of quantum circuits compared to classic computers \cite{Nature398:1999,ReportsProgPhys80:2017,Quantum_Inf_Process:2009, PRL:107:2011,Nature474:2011, NPJQuantum6:2020, AppliesPhysics:2019}. A particular class of SC is based on the so-called Josephson junctions~\cite{Nature474:2011,Devoret1169,Kockum2019, ReportsProgPhys80:2017,strambini2020josephson,santos2019stable,pal2019quantized}. These systems behave like artificial atoms~\cite{Nature474:2011} and have the advantage of presenting low dissipation, which yields long coherence times~\cite{Devoret1169}. In this regard, they can generate entanglement~\cite{Felicetti:Science:Reports:2017} and implement quantum gates~\cite{Zinner:Science:Reports:2019}, leading to the development of quantum hardware~\cite{Andrei:NaturePhysics:2020} with several applications in quantum information processing~\cite{Kockum2019,ReportsProgPhys80:2017,strambini2020josephson,santos2019stable,pal2019quantized,Felicetti:Science:Reports:2017,Zinner:Science:Reports:2019,Andrei:NaturePhysics:2020}.

In this work, we report the use of a superconducting quantum interference device composed of two Josephson junctions, the dc SQUID~\cite{Quantum_Inf_Process:2009, PhysRevLett.93.187003}, as a viable platform for generating a pair of twin photons and transfer quantum information. The generation of the photon pair takes place through the coupling between the superconducting device, which can be seen as an artificial atom, and a single-mode incident radiation field. This coupling yields entanglement dynamics emerging from the population transfer between the energy levels of the artificial atom, acting as our dc SQUID, leading to the generation of the twin photons. For the quantum information transference, we consider the dc SQUID initially in a maximally entangled state. The system absorbs the photon from the single-mode radiation field, and the modes of the photons generated by the population transfer will be entangled. As expected, we also observe that entanglement is transferred as the quantum coherence is transmitted from the SQUID to the modes since both properties, entanglement and coherence, stems from the quantum superposition principle, encapsulating the quantumness of the process. Our results reinforce the applicability of superconducting devices in the information transfer processes, contributing to the study of emerging quantum technologies based on quantum information transference.
\section{The dc SQUID}
\label{sec:squid_review}
The dc SQUID is a sensitive magnetic flux sensor composed of two Josephson junctions  placed in parallel in a quantum circuit with a critical current of intensity $I_{0}$, capacitance $C_{0}$, and phase differences $\phi_{1}$ and $\phi_{2}$, embedded in a loop of total inductance $L$ \cite{Quantum_Inf_Process:2009, PhysRevLett.93.187003}. Such junctions  act as nonlinear circuit elements, which ensures an unequal spacing between the energy levels~\cite{Nature474:2011}. Consequently, from an operational viewpoint, this superconducting circuit behave as quantum systems similar to an artificial atom with two degrees of freedom~\cite{PRL:107:2011,Andrei:NaturePhysics:2020,Nature474:2011}. However, unlike atoms, the coupling with the radiation field can be incorporated into the design of this device. In this circuit, Josephson's energy is given by $E = (\Phi_{0}/2\pi)I_{0}$, where $\Phi_{0}$ is the superconducting quantum flow \cite{Quantum_Inf_Process:2009}.

The phase dynamics of this system  can be understood as we map the evolution of a fictitious mass $m$ subjected to the two-dimensional potential $U(x,y)$, where $m$ and $U(x,y)$ depend on the electrical parameters of the superconducting device, being $x$ and $y$ degrees of freedom, $x=\left(\phi_{1}+\phi_{2} \right)$ and $y=\left(\phi_{1}-\phi_{2} \right)$. The Longitudinal (LM) and Transverse (TM) oscillation of this fictitious mass correspond to the parallel $(\Vert)$ and perpendicular $(\bot)$ modes. The nonlinear coupling of these modes leads to the emergence of promising quantum effects~\cite{PRB_83:2011,PRL:107:2011}, such as the creation of entanglement and the implementation of quantum logic gates~\cite{PRL:107:2011}. In this regard, we consider the coupling term of the total Hamiltonian of the system as a perturbation that acts on free Hamiltonian, and the interaction picture is used to determine the effective Hamiltonian, its energy eigenvalues $E_{n_{\Vert}, n_{\bot}}$ and the respective eigenstates $|n_{\Vert}\rangle\otimes|n_{\bot}\rangle$ of the coupled system. 

\begin{figure}
    \centering
    \includegraphics[scale=0.32]{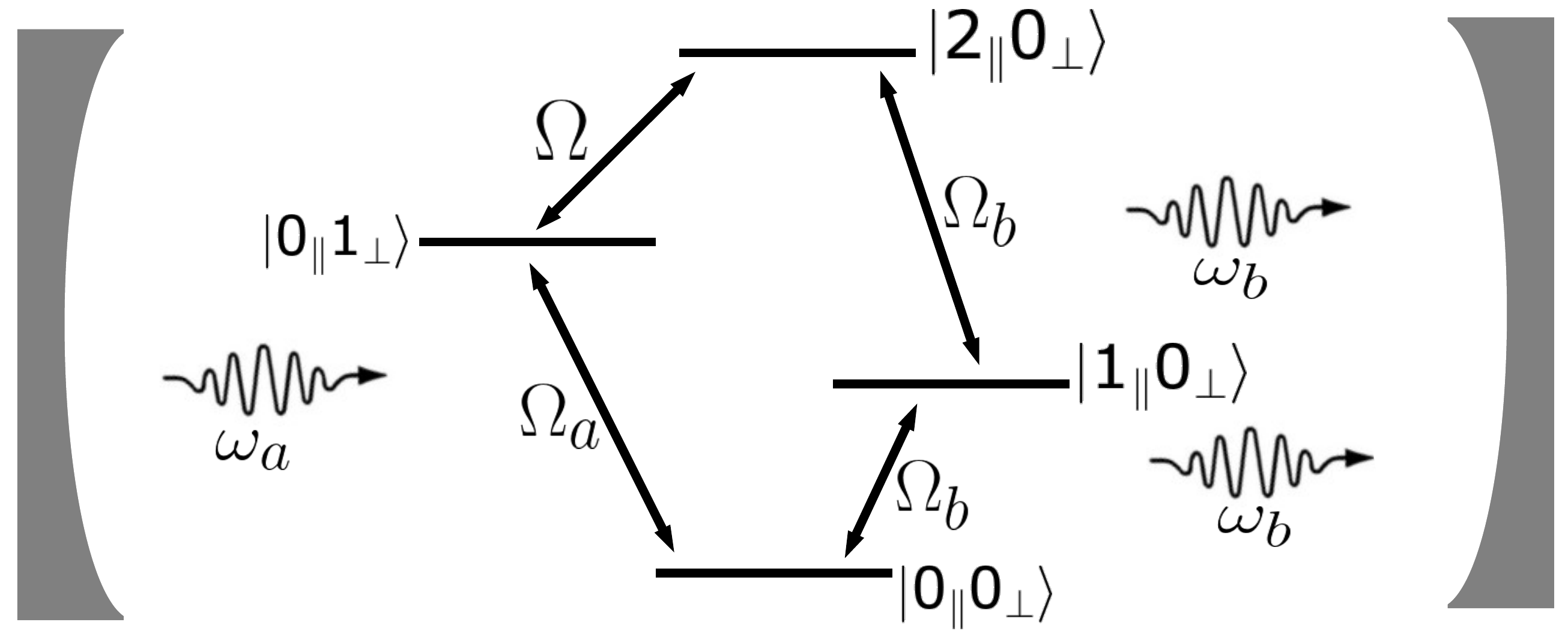}
    \caption{The first four energy levels of the dc SQUID. The superconducting device, whose population is initially at level $|0_{\Vert}0_{\perp}\rangle$, absorbs an incident photon in mode $\hat{a}$. As described in section III, a pair of photons in the mode $\hat{b}$ will be generated through the population transfer among the energy levels of the dc SQUID.}
    \label{fig:setup-squid}
\end{figure}

Considering the dc SQUID inserted in a microwave guide coupled to its modes and the coupling between the LM and  TM modes (see Fig.~\ref{fig:setup-squid}), the dynamics of this system is ruled by the Hamiltonian $\mathcal{H}=\mathcal{H}_{\text{free}} + \mathcal{H}_{\text{coupling}}$, where 
\begin{equation}    
\mathcal{H}_{\text{free}} =\sum_{i} \hbar \omega_{j} |j\rangle\langle j| + \hbar \omega_{a}\hat{a}^{\dagger}\hat{a} + \hbar \omega_{b}\hat{b}^{\dagger}\hat{b}
\label{eq:hamiltonian_free}
\end{equation} 
is the free Hamiltonian, with $\omega_{j}$ being the frequency of each energy level of the dc SQUID, represented by $|j\rangle$, and $\omega_{a(b)}$ the frequency of incident (emitted) photon; $\mathcal{H}_{\text{coupling}}$ is the Hamiltonian that couples the radiation field and the energy levels, according to the experimental data set out in ref. ~\cite{PRL:107:2011}. 

\begin{figure}
    \centering
    \includegraphics[scale=0.34]{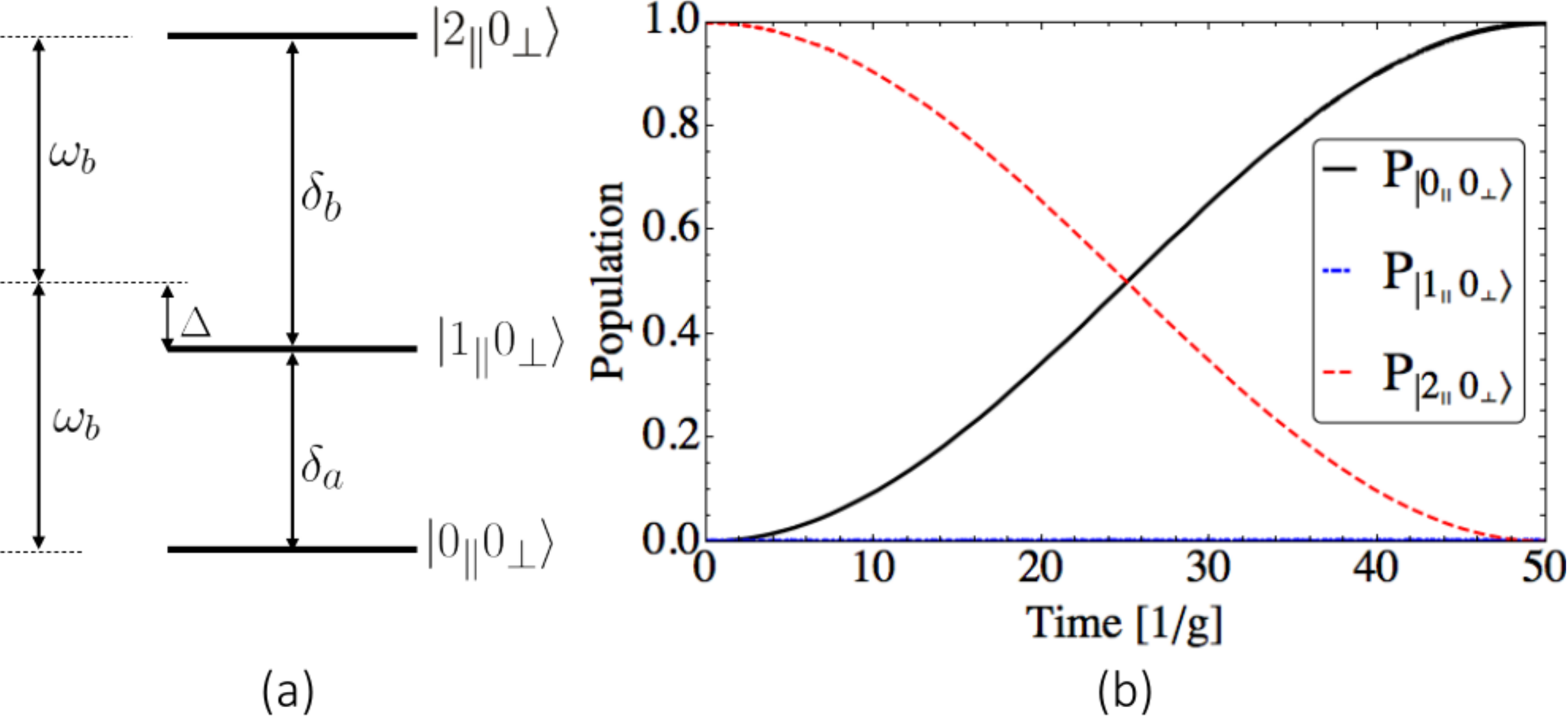}
    \caption{(color online). (a) $\Xi$-like structure level formed by energy eigenstates $|2_{\Vert}0_{\perp}\rangle$, $|1_{\Vert}0_{\perp}\rangle$ and $|0_{\Vert}0_{\perp}\rangle$. The energy gap between $|0_{\Vert}0_{\perp}\rangle$ and $|1_{\Vert}0_{\perp}\rangle$ is $\delta_{a}$, and the gap between $|2_{\Vert}0_{\perp}\rangle$ and $|1_{\Vert}0_{\perp}\rangle$ is given by $\delta_{b}$. The frequency of the incident photon is $\omega$ and $\Delta$ is energy difference between the SQUID energy levels and the incident photon. (b) Population transfer between the levels $|2_{\Vert}0_{\perp}\rangle$, $|1_{\Vert}0_{\perp}\rangle$, and $|0_{\Vert}0_{\perp}\rangle$. We consider the level $|2_{\Vert}0_{\perp}\rangle$ initially populated. As can be seen, the population is gradually  transferred to the level $|0_{\Vert}0_{\perp}\rangle$, without going through level $|1_{\Vert}0_{\perp}\rangle$.}
    \label{fig:three-level}
\end{figure}

The full coupling Hamiltonian is given by $\mathcal{H}_{\text{coupling}} = H_{1} + H_{2} + H_{3}$.
The first term $H_{1} = \hbar \Omega_{a}\left( \sigma_{a}^{+}\hat{a} + \sigma_{a}^{-}\hat{a}^{\dagger}  \right)$ refers to coupling between the energy levels $|0_{\Vert}0_{\perp}\rangle$ and $|0_{\Vert}1_{\perp}\rangle$ with an incident radiation field in mode $\hat{a}$, where $\sigma_{a}^{+} = |0_{\Vert}1_{\perp}\rangle\langle 0_{\vert}0_{\perp}|$ and $\sigma_{a}^{-} = |0_{\Vert}0_{\perp}\rangle\langle 0_{\Vert}1_{\perp}|$ are the raising and lowering atomic operators respectively. The second term  $H_{2} = \hbar \Omega\left( \sigma^{+} + \sigma^{-} \right)$ describes the couple between the levels $|0_{\Vert}1_{\perp}\rangle$ and $|2_{\Vert}0_{\perp}\rangle$, where $\sigma^{+} = |2_{\Vert}0_{\perp}\rangle\langle 0_{\Vert}1_{\perp}|$ and $\sigma^{-} = |0_{\Vert}1_{\perp}\rangle\langle 2_{\Vert}0_{\perp}|$ are the ladder operators, with $\Omega$ being the coupling constant. Finally, the third term ($H_{3}$) refers to the coupling between the another three levels, $|2_{\vert}0_{\perp}\rangle$, $|1_{\Vert}0_{\perp}\rangle$ and $|0_{\Vert}0_{\perp}\rangle$, which form a $\Xi$-like structure level (see Fig.~\ref{fig:three-level}).  The gap between the energy levels $|0_{\Vert}0_{\perp}\rangle$ and $|1_{\vert}0_{\perp}\rangle$ is $\delta_{a}$, while the gap between the levels $|2_{\Vert}0_{\perp}\rangle$ and $|1_{\Vert}0_{\perp}\rangle$ is $\delta_{b}$, with $\Delta = \hbar\omega - \delta_{a} = \delta_{b} - \hbar\omega$ being the energy difference between the energy levels and the incident photon. Using the interaction picture we obtain the effective Hamiltonian for this subsystem as 
\begin{equation}
H_{3} = \hbar\frac{g^2}{\Delta} \left[ b^{2}|2_{\Vert}0_{\perp}\rangle\langle 0_{\Vert}0_{\perp}| + b^{\dagger 2}|0_{\Vert}0_{\perp}\rangle\langle 2_{\Vert}0_{\perp}| \right],
\end{equation}
where $\hbar g$ is the coupling strength among the energy levels.

Fig.~\ref{fig:three-level}(b) shows the population transfer among the energy levels, $P_{|2_{\Vert}0_{\perp}\rangle}$, $P_{|1_{\Vert}0_{\perp}\rangle}$ and $P_{|0_{\Vert}0_{\perp}\rangle}$, considering the energy level $|2_{\vert}0_{\perp}\rangle$ initially populated and $g/\Delta \approx 10^{-2}$. Furthermore, for simplicity we take $\Omega_{b} \equiv g^2/\Delta$. Thus, the population is directly transferred to energy level $|0_{\Vert}0_{\perp}\rangle$, without going through energy level $|1_{\vert}0_{\perp}\rangle$. 
\section{Population Transfer}

In order to transfer population among the levels of the superconducting device, we consider that the energy level $\vert 0_{\Vert}1_{\bot}\rangle$ moves up, as its energy increases over time, while the level $\vert 2_{\Vert}0_{\bot}\rangle$ moves down as its energy decreases. Thus, we can define the frequencies 
\begin{eqnarray}
\omega_{0_{\Vert}1_{\bot}}(t) &=& \omega_{0_{\Vert}1_{\bot}}(0) \left( 1 + v_{1} t \right), \label{dinamic1}\\
\omega_{2_{\Vert}0_{\bot}}(t) &=& \omega_{2_{\Vert}0_{\bot}}(0) \left( 1 - v_{2}t \right), \label{dinamic2}
\end{eqnarray}
where $v_{1}$ and $v_{2}$ are the rates of variation of the energy of the levels $|0_{\Vert}1_{\bot}\rangle$  and $|2_{\Vert}0_{\bot}\rangle$ respectively, with $\hbar \omega_{n_{\Vert}n_{\bot}}(0)$ being the respective energies at instant $t=0$. Therefore, the system dynamics can be obtained from the interaction picture as the levels $|0_{\Vert}1_{\bot}\rangle$ and $|2_{\Vert}0_{\bot}\rangle$, coupled by the constant $\Omega$, move in opposite directions with equal rates {$v_{1} = v_{2}$}. 

In the interaction picture, the interaction Hamiltonian is given by
\begin{equation}
\mathcal{H}_{\text{int}} = \mathcal{U}(t) \mathcal{H}_{\text{coupling}} \mathcal{U}^{\dagger}(t)
\label{eq:hamiltonian_interaction}
\end{equation}
where
\begin{equation}
\mathcal{U}(t) = \exp{i \int_{0}^{t} dt^{\prime} \mathcal{H}_{\text{free}}(t^{\prime})}.
\label{U}
\end{equation}
The result of this integral is 
\begin{eqnarray}
\int_{0}^{t} dt^{\prime} H_{\text{free}}(t^{\prime})  &=& \omega_{0_{\Vert}0_{\bot}} t |0_{\Vert}0_{\bot}\rangle \langle 0_{\Vert}0_{\bot}| + r_{0_{\Vert}1_{\bot}}(t) |0_{\Vert}1_{\bot}\rangle \langle 0_{\Vert}1_{\bot}|+ \nonumber \\ &&  +  \omega_{a} t \hat{a}^{\dagger}\hat{a} + r_{2_{\Vert}0_{\bot}}(t) |2_{\Vert}0_{\bot}\rangle \langle 2_{\Vert}0_{\bot}| +\omega_{b}t \hat{b}^{\dagger} \hat{b},
\label{intH0}
\end{eqnarray}
where
\begin{eqnarray}
r_{0_{\Vert}1_{\bot}}(t) &=& \omega_{0_{\Vert}1_{\bot}}(0) \left( t + \frac{1}{2} v_{1} t^{2} \right), \\
r_{2_{\Vert}0_{\bot}}(t) &=& \omega_{2_{\Vert}01_{\bot}}(0) \left( t - \frac{1}{2} v_{2} t^{2} \right).
\end{eqnarray}

In this context, the dynamics of this system is ruled by the time-dependent Schr\"{o}dinger equation into the interaction picture
\begin{equation}
i \hbar \frac{d}{dt} |\psi_{\text{int}}(t)\rangle = \mathcal{H}_{\text{int}} |\psi_{\text{int}}(t)\rangle,
\label{eq:interaction_picture}
\end{equation}
where
\begin{eqnarray}
|\psi_{\text{int}}(t)\rangle &=& c_{1}(t) |1\rangle_{a}|0\rangle_{b}|0_{\Vert}0_{\bot}\rangle + c_{2}(t)|0\rangle_{a}|0\rangle_{b}|0_{\Vert}1_{\bot}\rangle + \nonumber \\
&+& c_{3}(t)|0\rangle_{a}|0\rangle_{b}|2_{\Vert}0_{\bot}\rangle + c_{4}(t)|0\rangle_{a}|2\rangle_{b}|0_{\Vert}0_{\bot}\rangle~.
\label{psi}
\end{eqnarray}
The coefficient $\vert c_{k}(t)\vert^{2}$ provides the occupation  probability $P_{k}$ of the respective state coupled by $\mathcal{H}_{\text{int}}$.

Applying Eq. (\ref{psi}) in Eq. (\ref{eq:interaction_picture}) we obtain a set of coupled differential equations for the $ c_{k}(t)$ coefficients
\begin{widetext}
\begin{eqnarray}
    \dot{c}_{1}(t) &=&-i\lbrace\Omega_{a}e^{i\left[(\omega_{0_{\Vert}1_{\bot}} + \omega_{a})t -ir_{0_{\Vert}1_{\bot}}\right]}c_{2}(t)\rbrace~,
    \label{c1} \\
    \dot{c}_{2}(t)&=&-i\lbrace\Omega_{a}e^{i\left[ r_{0_{\Vert}1_{\bot}} - (\omega_{0_{\Vert}1_{\bot}} + \omega_{a})t \right]}c_{1}(t)+ \Omega e^{i\left[r_{0_{\Vert}1_{\bot}} - r_{2_{\Vert}0_{\bot}}(t) \right]}c_{3}(t)\rbrace~,
    \label{c2} \\
    \dot{c}_{3}(t)&=&-i\lbrace\Omega e^{i\left[r_{2_{\Vert}0_{\bot}}(t) - ir_{0_{\Vert}1_{\bot}}\right]}c_{2}(t)+ \sqrt{2}\Omega_{b}e^{i\left[r_{2_{\Vert}0_{\bot}}(t) - (\omega_{0_{\Vert}1_{\bot}}  + 2\omega_{b})t\right]}c_{4}(t)\rbrace ~,
    \label{c3} \\
    \dot{c}_{4}(t)&=&-i\lbrace \sqrt{2}\Omega_{b}e^{i\left[(\omega_{0_{\Vert}1_{\bot}}  + 2\omega_{b})t - r_{2_{\Vert}0_{\bot}}(t)\right] }c_{3}(t)\rbrace~.
    \label{c4}
\end{eqnarray}
\end{widetext}
The solution of these coupled differential equations, Eqs. (\ref{c1}) - (\ref{c4}), gives us the dynamics of the population of the system. In order to show the generation of two photons using a dc SQUID, we consider the radiation field incident on dc SQUID as a single photon in the mode $\hat{a}$ and frequency $\omega_{a}$, as the initial condition to Eqs. (\ref{c1}) - (\ref{c4}).  The initial state is given by \begin{equation}
{
|\psi_{0}\rangle = |1\rangle_{a}|0\rangle_{b}|0_{\Vert}0_{\bot}\rangle~.}
\label{psi_single}
\end{equation}
This state represents one photon in the mode $\hat{a}$, the initial population of the superconducting device in the energy level $E_{0_{\Vert}0_{\bot}}$, and no photon in the mode $\hat{b}$. In the following, the superconducting {device} absorbs an incident photon and the population is transferred to level  $|0_{\Vert}1_{\bot}\rangle$. The coupling strength between these levels ($10^{-2} \Omega$) has the same order of magnitude of the coupling between levels between $|0_{\Vert}0_{\bot}\rangle$ and $|2_{\Vert}0_{\bot}\rangle$. Consequently, the population will be transferred from level $|0_{\Vert}1_{\bot}\rangle$ to $|2_{\Vert}0_{\bot}\rangle$. Thus, the population is transferred from level $|2_{\Vert}0_{\bot}\rangle$ to level $|0_{\Vert}0_{\bot}\rangle$ and a de-excitation occurs creating a pair of photons in the mode $\hat{b}$. Fig. \ref{fig:setup-squid} shows a schematic representation of this dynamic process.

\begin{figure}[!h]
    \centering
    \includegraphics[scale=0.45]{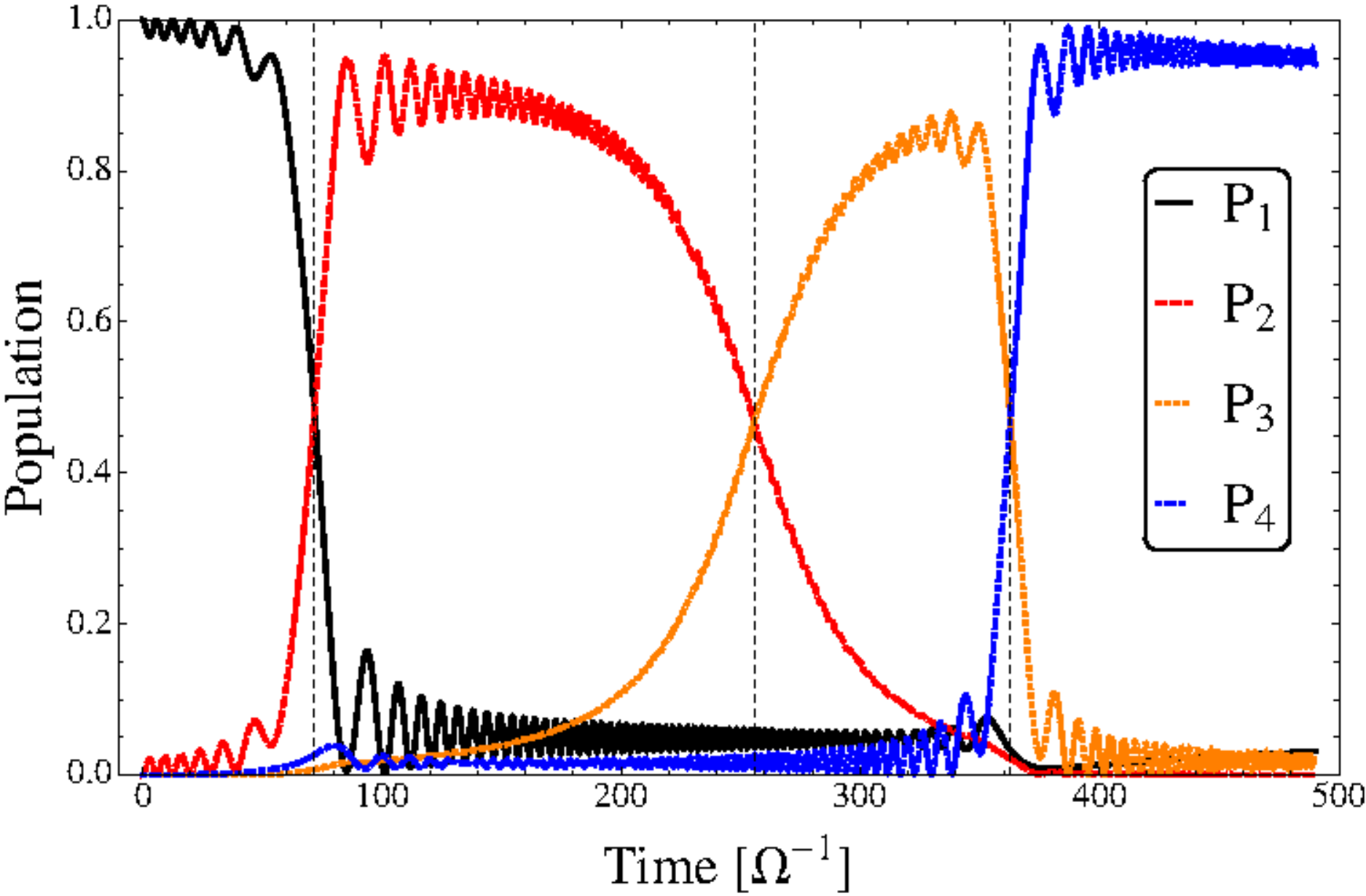}
    \caption{(Color online) Generation of a pair of photons in the mode $\hat{b}$ using a single mode as incident field, according to the realistic experimental parameters obtained from Ref.~\cite{PRL:107:2011}. The black line (P$_{1}$) labels the state  $|1\rangle_{a}|0\rangle_{b}|0_{\vert}0_{\bot}\rangle$; the red line (P$_{2}$) labels {the state} $|0\rangle_{a}|0\rangle_{b}|0_{\Vert}1_{\bot}\rangle$; the orange line (P$_{3}$) labels the state $|0\rangle_{a}|0\rangle_{b}|2_{\Vert}0_{\bot}\rangle$; the blue line (P$_{4}$) labels the state $|0\rangle_{a}|2\rangle_{b}|0_{\Vert}0_{\bot}\rangle$. The vertical dashed lines indicate the time when the population curves of each state cross. The time is in units of  $\Omega^{-1}$, where $\Omega$ is the coupling constant between the modes $|0_{\Vert}1_{\bot}\rangle$ and $|2_{\Vert}0_{\bot}\rangle$. }
    \label{fig:transfer_single_mode}
\end{figure}

Fig.~\ref{fig:transfer_single_mode} shows the numerical solution for the population of the states $|n_{a}\rangle\otimes|n_{b}\rangle\otimes|n_{\Vert}n_{\bot}\rangle$ in terms of coupling constant $\Omega$, obtained from  Eqs. (\ref{c1}) - (\ref{c4}), where $|n_{a}\rangle$ are the incident radiation field states (incident photon), $|n_{b}\rangle$ are the generated radiation field states (generated photons), and $|n_{\Vert}n_{\bot}\rangle$ are the eigenstates of the coupling Hamiltonian, with time in units of  $\Omega^{-1}$. We use realistic experimental parameters from Ref.~\cite{PRL:107:2011} in this process. 

As can be seen in Fig.~\ref{fig:transfer_single_mode}, the dynamic process of generation of the photons can be divided in four main steps: 

i. Firstly, the system is at state $|1\rangle_{a}|0\rangle_{b}|0_{\Vert}0_{\bot}\rangle$, which means that energy level  $E_{0_{\Vert}0_{\bot}}$ is completely populated. When the photon in mode $\hat{a}$ is absorbed by dc SQUID, the population will be transferred to the level $|0_{\Vert}1_{\bot}\rangle$, {leading to a crossing of the (solid) black and (dashed) red curves, that represent the population transfer between the energy levels $E_{0_{\Vert}0_{\bot}}$ and $E_{0_{\Vert}1_{\bot}}$ of the superconducting device.}

ii. In the following, due to coupling between the energy levels $E_{0_{\Vert}1_{\bot}}$ and $E_{2_{\Vert}0_{\bot}}$, the population will  be transferred from the state $|0\rangle_{a}|0\rangle_{b}|0_{\Vert}1_{\bot}\rangle$ to $|0\rangle_{a}|0\rangle_{b}|2_{\Vert}0_{\bot}\rangle$, leading to a crossing between the {(dashed) red  and (dotted) orange curves, that represent the populations in the energy levels $E_{0_{\Vert}1_{\bot}}$ and $E_{2_{\Vert}0_{\bot}}$, respectively.}

iii. Then,  due to the de-excitation, the population of the level $E_{2_{\Vert}0_{\bot}}$ begins to be transferred to back to the level $E_{0_{\Vert}0_{\bot}}$ and, as a consequence, two photons will be generated in mode $\hat{b}$, leading to the crossing {of the (dotted) orange  and  blue population curves, that represent, respectively,} the states $|0\rangle_{a}|0\rangle_{b}|2_{\Vert}0_{\bot}\rangle$ and $|0\rangle_{a}|2\rangle_{b}|0_{\Vert}0_{\bot}\rangle$, respectively..

iv. Finally, this process ends with the population returning to initial energy level of SQUID and a pair of photons is generated.

Therefore, the system initiates the process in the state $|1\rangle_{a}|0\rangle_{b}|0_{\Vert}0_{\bot}\rangle$, Eq. (\ref{psi_single}), after the above described dynamic process the population is transferred to the final state $|0\rangle_{a}|2\rangle_{b}|0_{\Vert}0_{\bot}\rangle$, with two photons in the mode $b$, where each photon created has half the frequency of the radiation field in mode $\hat{a}$. However, we cannot say that this process is parametric down conversion because we cannot derive an effective Hamiltonian that describes this type of process \cite{Couteau:2018}. 

\subsection{Entanglement Dynamics}

On the other hand, the population crossover observed in Fig.~\ref{fig:transfer_single_mode} can be understood in terms of an entanglement dynamics between the SQUID and the photons, incident and generated. We quantify entanglement between the SQUID and the photons by using Entanglement of Formation \cite{wootters,hill,horodecki}, which is defined as
\begin{equation}
\mathbb{E}=-\mathbb{E}_+-\mathbb{E}_-
\label{eq:entanglement}
\end{equation}
where 
\begin{equation}
\mathbb{E}_\pm=\frac{1\pm\sqrt{1-\mathbb{C}^2}}{2}\log_2 \left(\frac{1\pm\sqrt{1-\mathbb{C}^2}}{2}\right)
\end{equation}
and $\mathbb{C}=\max[0, \sqrt{\lambda_1} - \sqrt{\lambda_2} - \sqrt{\lambda_3}- \sqrt{\lambda_4}]$ is the concurrence~\cite{wootters,hill,horodecki}, where $\lambda_1 \geq \lambda_2 \geq \lambda_3 \geq \lambda_4 \geq 0$ are the eigenvalues of the matrix $R(t) = \rho(t) (\sigma_y \otimes \sigma_y) \rho^*(t) (\sigma_y \otimes \sigma_y)$ \cite{wootters,hill,horodecki}, with $\sigma_y$ being the y-Pauli matrix and $\rho(t) = |\psi_{\text{int}} (t)\rangle\langle \psi_{\text{int}} (t)|$.

\begin{figure}
    \centering
    \includegraphics[scale=0.45]{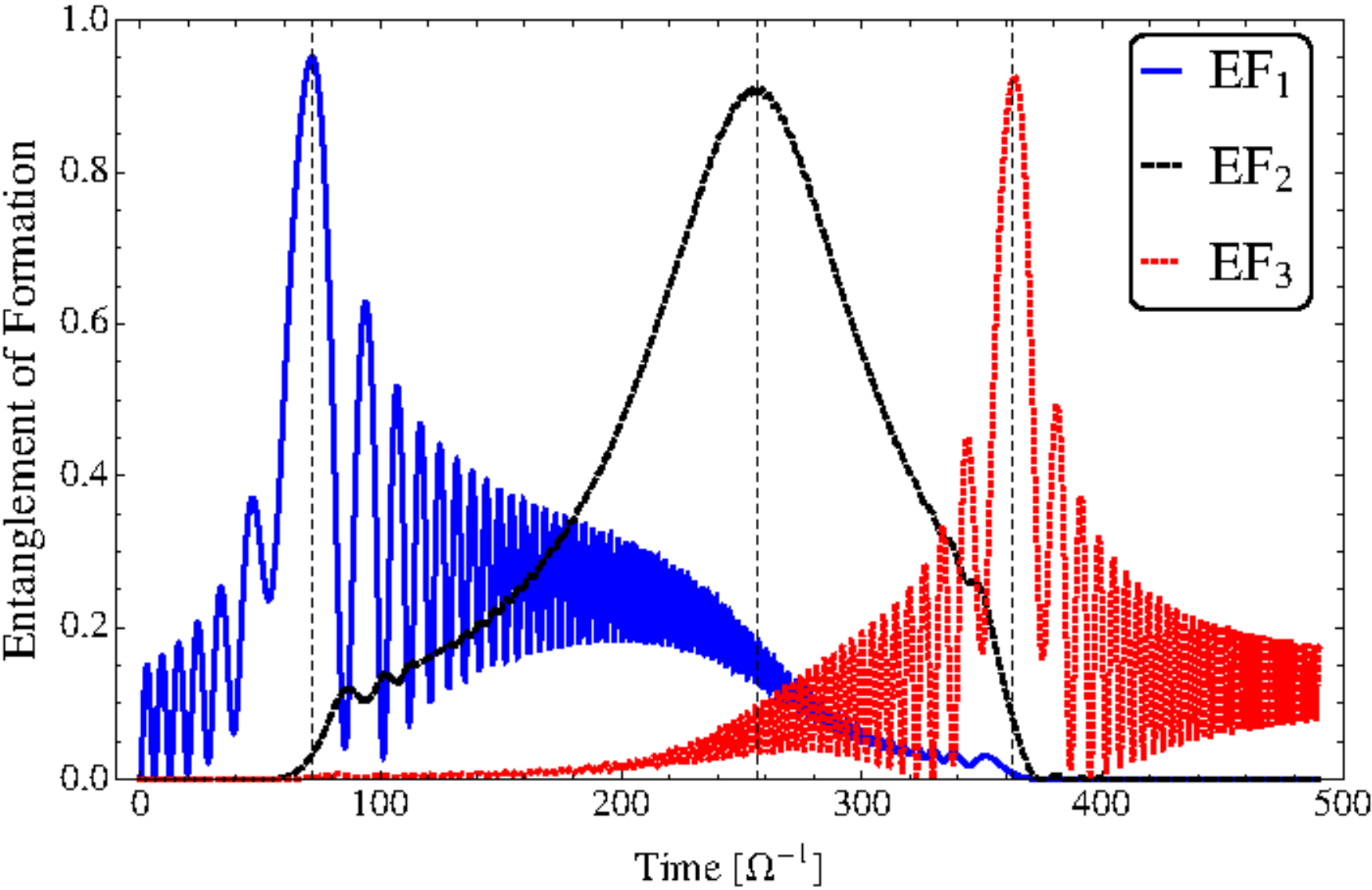}
    \caption{(Color online) Entanglement of Formation ($EF$) for each crossing {among} the populations presented in Fig.~\ref{fig:transfer_single_mode}. The (solid) blue line ($EF_{1}$) refers to the entanglement between the incident radiation field and the superconducting device; the (dashed) black line ($EF_{2}$) refers to an intra-SQUID entanglement; and, finally, the (doted) red line ($EF_{3}$) refers to entanglement between the dc SQUID and the pair of generated photons. It is worth noting that $EF$ is maximum in each crossing point between the curves in Fig.~\ref{fig:transfer_single_mode}.}
    \label{fig:entanglement_populations}
\end{figure}

Fig.~\ref{fig:entanglement_populations} shows entanglement dynamics for each population crossover showed in Fig.~\ref{fig:transfer_single_mode}. As can be seen, the incident radiation field in mode $\hat{a}$ entangles with the SQUID as the population is transferred from level $|0_{\Vert}0_{\bot}\rangle$ to $|0_{\Vert}1_{\bot}\rangle$; in the following, the degree of entanglement between the incident field {and the dc SQUID} decreases, while the intra-SQUID entanglement increases along with the entanglement between the dc SQUID and the pair of generated photons; thereafter, when the intra-SQUID entanglement vanishes, the entanglement between the {superconducting} device and the pair of generated photons is maximum. Finally, as the population is transferred from the energy level $|2_{\Vert}0_{\bot}\rangle$ to  $|0_{\Vert}0_{\bot}\rangle$, a pair of photons is generated and the degree of entanglement between these photons and the SQUID decreases up to a minimal value. Thus, the pair of generated photons will remain entangled with the superconducting device at a minimum degree. Therefore, the population transfer happened due to an entanglement dynamics between SQUID and photons, incident and generated, reinforcing the role of superconducting devices on the quantum information transference processes.

\section{Transference of Quantum Entanglement and Coherence}

As presented in the last section, the dc SQUID can be a useful tool for the information transfer processes. In order to study the role of this superconducting circuit in the transference of quantum information, we will explore the dynamics of transference of quantum entanglement from the dc SQUID to the radiation field modes, $\hat{a}$ and $\hat{b}$. 
Let us consider an entangled dc SQUID, where we populate equally the levels $|0_{\Vert}1_{\bot}\rangle$ and $|2_{\Vert}0_{\bot}\rangle$ without any photon in the modes $\hat{a}$ or $\hat{b}$, as the initial condition to Eqs. (\ref{c1}) - (\ref{c4}). Thus, we can write the  maximally entangled initial state of the system as 
\begin{eqnarray}
    |\psi_{0}\rangle = \frac{1}{\sqrt{2}}\left( |0\rangle_{a}|0\rangle_{b}\right) \left( |0_{\Vert}1_{\bot}\rangle + |2_{\Vert}0_{\bot}\rangle \right).
    \label{psi_entangled}
\end{eqnarray}

Considering the same dynamics described in Eqs. (\ref{dinamic1}) and (\ref{dinamic1}), with rate $v_{1} = 2v_{2}$, in order to speed up the population transference, a radiation field in the single mode $\hat{a}$ will affect the dc SQUID, leading to the complete transference of the population from the device to the photons in the modes $\hat{a}$ and $\hat{b}$. In other words, the entanglement initially on the dc SQUID, Eq. (\ref{psi_entangled}), will be completely transferred to the  modes, with final state:
\begin{equation}
    |\psi\rangle = \frac{1}{\sqrt{2}}\left( |1\rangle_{a}|0\rangle_{b}+ |0\rangle_{a}|2\rangle_{b}\right)\left(  |0_{\Vert}0_{\bot}\rangle \right).
    \label{finalstate}
\end{equation}

Therefore, there is a transference of entanglement from the superconducting device to the generated radiation field modes. In order to examine the entanglement dynamics through the transference process, we need to measure the entanglement of formation in the SQUID and the modes separately. From the partial trace of the interaction density matrix $\rho(t) = |\psi_{\text{int}} (t)\rangle\langle \psi_{\text{int}} (t)|$ for each subsystem, we obtain the modes density matrix as
\begin{eqnarray}
\rho_{ab}(t) &=& \text{Tr}_{SQUID}\left[\rho(t)]\right] \nonumber \\
&=& \left(
\begin{array}{cccc}
|c_{2}(t)|^{2} + |c_{3}(t)|^{2} & 0 & 0 & 0 \\
0 & |c_{1}(t)|^{2} & c_{1}^{*}(t) c_{4}(t) & 0 \\
0 & c_{1}(t) c_{4}^{*}(t) & |c_{4}(t)|^{2} & 0 \\
0 & 0 & 0 & 0
\end{array}
\right)~,
\label{eq:rho_ab}
\end{eqnarray}
and the SQUID density matrix as
\begin{eqnarray}
\rho_{SQUID} &=& \text{Tr}_{ab} \left[\rho(t)\right] \nonumber \\
&=& \left(
\begin{array}{cccc}
|c_{1}(t)|^{2} + c_{4}(t)|^{2} & 0 & 0 & 0 \\
0 & |c_{2}(t)|^{2} & c_{2}(t) c_{3}^{*}(t) & 0 \\
0 & c_{2}^{*}(t) c_{3}(t) & |c_{3}(t)|^{2} & 0 \\
0 & 0 & 0 & 0
\end{array}
\right),
\label{eq:rho_squid}
\end{eqnarray}
using the reference basis $\{ |n_{a}\rangle\otimes|n_{b}\rangle\otimes|n_{\Vert}n_{\bot}\rangle \}$. From Eq.(\ref{eq:entanglement}), re obtain the entanglement dynamics for the SQUID and the modes in terms of the probability amplitudes $|c_{k}(t)|^{2}$.

\begin{figure}[!h]
    \centering
    {\includegraphics[scale=0.38]{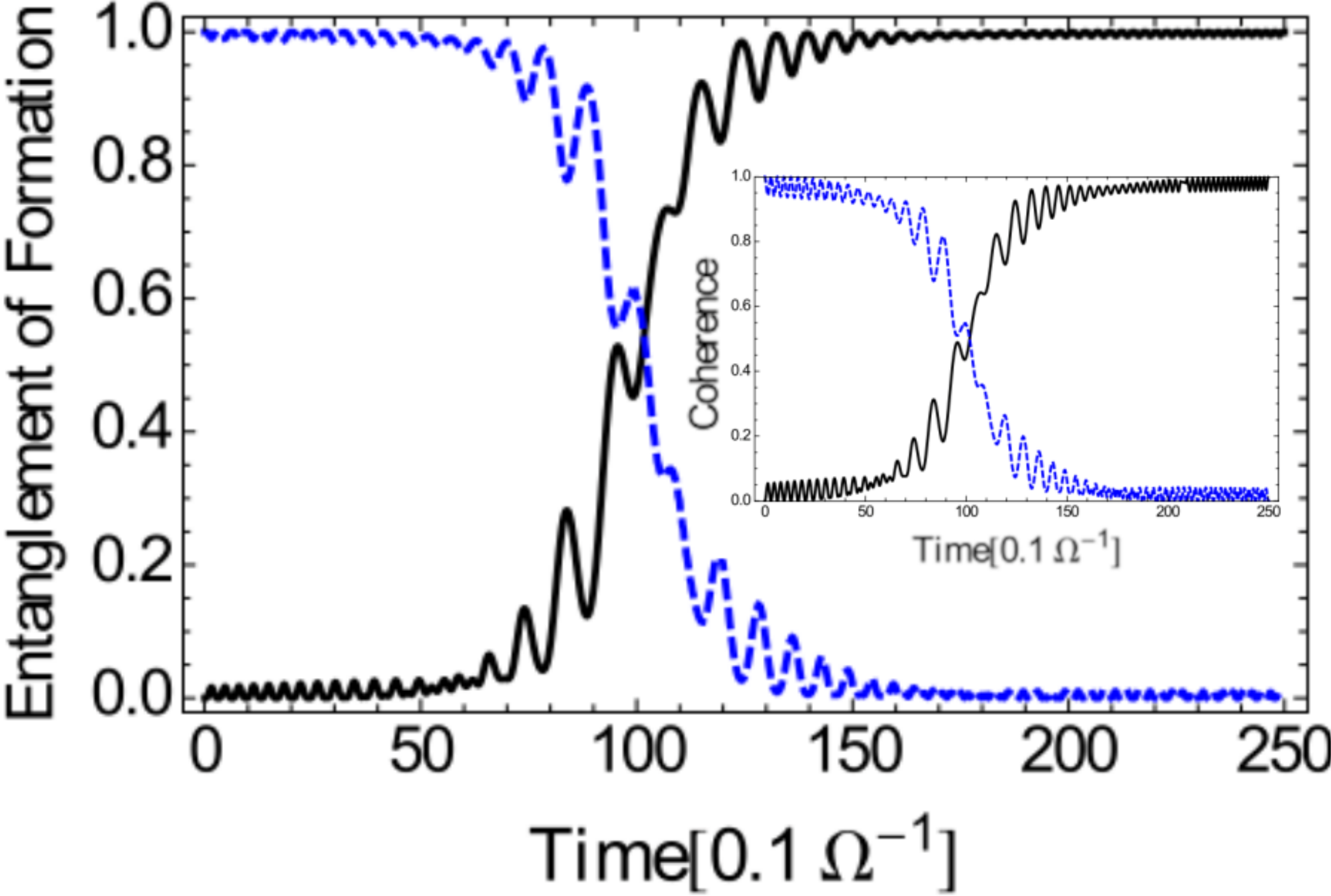}}
    \caption{(Color online) Entanglement and Coherence (inset) transference from the SQUID (dashed blue line) to the field modes $a$ and $b$ (solid black line), according to the realistic experimental parameters obtained from Ref.~\cite{PRL:107:2011}. The initial and final state are given by Eqs.~(\ref{psi_entangled}) and (\ref{finalstate}), respectively. Levels $|1_{\Vert}0_{\bot}\rangle$ and $|2_{\Vert}0_{\bot}\rangle$ are initially equally populated and move with rates $v_{1}$ and $v_{1}$, respectively, where $v_{1} = 2v_{2}$ was selected to speed up this transference process.
    As can expected, the entanglement is transferred form the SQUID to the modes as the quantum coherence is transferred, since both stems from the quantum superposition principle, encapsulating the quantumness of the process.}
    \label{fig:entanglement}
\end{figure}

Fig. \ref{fig:entanglement} shows the {entanglement transfer} {from} the dc SQUID to the modes $\hat{a}$ and $\hat{b}$. As can be seen, the  entanglement is initially in the SQUID and a radiation field in the single mode $\hat{a}$ affects {it} causing transfer of the entanglement to the radiation field modes. This result shows that entanglement can be  transferred  using  a superconducting device, reinforcing the fact that dc SQUID is an important tool for use in tasks in quantum information processing and it can be useful for information transference processes.

On the other hand, just as entanglement can be transferred (see Fig. \ref{fig:entanglement}), we can also explore the transfer of coherence between SQUID and the {radiation field} modes, since the quantum coherence is a necessary feature for different forms
of quantum correlations \cite{xi2015quantum,hu2018quantum,yadin2016quantum,egloff2018local,streltsov2015measuring}. Coherent superposition of quantum states embodies the nature of the entanglement, being a resource for several quantum processes in quantum optics, solid state physics, quantum game theory, quantum metrology and thermodynamics \cite{cruz2020quantifying,PhysRevLett.113.170401,giovannetti2011advances,lambert2013quantum,hu2018quantum,theurer2019quantifying,yadin2019coherence,xi2015quantum,streltsov2016quantum,kammerlander2016coherence,goold2016role,santos2020entanglement,passos2019non,yadin2016quantum,egloff2018local,streltsov2015measuring,lostaglio2019introductory,biswas2017interferometric}
Baumgratz \textit{et al.} \cite{baumgratz2014quantifying} defined a consistent theoretical basis to quantifies quantum coherence in a quantum state $\rho$ \cite{streltsov2016quantum,hu2018quantum}. From a geometric approach, is possible to measure the quantum coherence as:
\begin{eqnarray}
\mathcal{C}_D=\min_{\lbrace \sigma \in \mathcal{I}\rbrace} D(\rho,\sigma),
\end{eqnarray}
where $D(\rho,\sigma)$ is the distance measurement, between the state of interest $\rho$ and a set of incoherent states $\lbrace \sigma=\sum_{k}^{d} \vert k\rangle\langle k \vert \in \mathcal{I} \rbrace$  in a $d$-dimensional Hilbert space. Using $l_{1}$ trace norm as the distance measurement \cite{streltsov2016quantum,baumgratz2014quantifying,hu2018quantum,rana2016trace}, $l_{1}$ trace norm quantum coherence can be written as
\begin{eqnarray}
\mathcal{C}_{l_{1}}&=& \min_{\sigma \in \mathcal{I}} \Vert \rho -\sigma \Vert_{l_1}=\sum_{m\neq n} \vert \langle m\vert \rho\vert n \rangle\vert~.
\label{eq:coherence}
\end{eqnarray}

From Eq.~(\ref{eq:coherence}), we obtain the quantum coherence in the reference basis $\{ |n_{a}\rangle\otimes|n_{b}\rangle\otimes|n_{\Vert}n_{\bot}\rangle \}$ for the modes and the SQUID. The inset of Fig. \ref{fig:entanglement} shows the  $l_{1}$ trace norm quantum coherence transferred from the SQUID to modes $\hat{a}$ and $\hat{b}$. The initial and final states are given by Eqs.~(\ref{psi_entangled}) and (\ref{finalstate}), respectively. As can be seen, from Fig \ref{fig:entanglement} entanglement is transferred from  dc SQUID to {the radiation field} modes as the coherence is transferred as expected, since both entanglement and coherence stems from the quantum superposition principle, encapsulating the quantumness of the information transfer process.

\section{Conclusion}

In this work, we present the use of a dc SQUID as a feasible information transfer device for the generation of a pair of photons and  the transference of quantum information between the device and the radiation field modes.

We show that a pair of twin photons can be generated through the coupling between dc SQUID, described as an artificial atom, and a single-mode radiation field. We consider a single-mode incident radiation field, where energy levels $E_{0_{\Vert}1_{\bot}}$ and $E_{2_{\Vert}0_{\bot}}$ of the SQUID moves in opposite directions at the same rate. This process yields an entanglement dynamics, emerging from the population crossover between the energy levels of the superconducting device, which leads to the generation of the twin photons.

Moreover, we explore the transfer of quantum entanglement and coherence from dc SQUID to the radiation field modes. In this case, we considered the energy levels $E_{0_{\Vert}1_{\bot}}$ and $E_{2_{\Vert}0_{\bot}}$ equally populated, so dc SQUID initiates the dynamic process maximally entangled. Again, a radiation field in the single mode {$\hat{a}$} affects {the} superconducting device and, as a consequence, the radiation field modes becomes entangled as the quantum coherence is transferred from {the} dc SQUID to the radiation field modes. 

Therefore, our results shown the applicability of superconducting devices for the transference of quantum information, contributing to the study of emerging quantum technologies based on the information transfer process.
\begin{acknowledgments}
C. Castro gratefully acknowledges Mario Reis  and Perola Milman for the valuable discussions. This study was financed in part by the CNPq and  the \textit{Coordena\c{c}\~{a}o de Aperfei\c{c}oamento de Pessoal de N\'{i}vel Superior - Brasil} (CAPES) - Finance Code 001.
\end{acknowledgments}
%

\end{document}